\documentstyle[aps,prl,twocolumn,graphicx]{revtex}
\newcommand{\dn}[2]{{d}^{#1} #2 \:}
\newcommand{\dntwopi}[2]{\frac{{d}^{#1} #2}{(2\pi)^{#1}}\:}
\newcommand{\eqref}[1]{(\ref{#1})}

\newcommand{\qinv}{q_{\rm inv}}
\begin{document}
\draft
%-----------------------------------------------------------
%   Frontmatter
%
\wideabs{
\title{\begin{flushright}{\rm DOE/ER/40561-66-INT99}\\[9mm]\end{flushright}
Implications of the unusual structure in the pp correlation from 
Pb+Pb collisions at 158~AGeV}
%\title{Questioning the assumptions of two-proton correlations 
%in Pb+Pb collisions at 158~AGeV}

%\author{David A. Brown} 
%\email{dbrown@phys.washington.edu}
%\affiliation{University of Washington}
%\author{Fuqiang Wang}
%\affiliation{Lawrence Berkeley National Laboratory}
%\author{Pawe{\l} Danielewicz}
%\affiliation{National Superconducting Cyclotron Laboratory}

\author{David A. Brown$^1$, Fuqiang Wang$^2$, and Pawe{\l} Danielewicz$^3$}

\address{
$^1$ Institute for Nuclear Theory, University of Washington, Box 351550,
Seattle, WA 98195-1550, USA\\
%Electronic address: {\tt dbrown@phys.washington.edu}\\
$^2$ Nuclear Science Division, Lawrence Berkeley National Laboratory,
	Berkeley, CA 94720, USA\\
$^3$ National Superconducting Cyclotron Laboratory and Department of 
Physics and Astronomy, Michigan State University, East Lansing, MI 48824, USA}

\date{\today}
\maketitle
%-----------------------------------------------------------
%   Abstract
%
\begin{abstract}
The recent NA49 measurement of two-proton correlation function 
shows an interesting and unexpected structure 
at large relative momentum.
%in the tail.  
Applying source imaging techniques to the measurement,
we find an unusually steep drop-off in the two-proton source function.
We show that the steep drop-off is due to the structure in the 
correlation and the drop-off cannot
be explained using conventional correlation analysis.  
We suggest possible physics reasons for the unusual source function.
\end{abstract}

}
%
%===========================================================
%\section*{Introduction}

The NA49 collaboration has recently measured the two-proton 
correlation function from Pb+Pb collisions at 158 AGeV, integrated
over transverse momentum in the rapidity range $2.4<y<3.4$ 
(midrapidity $\approx 2.9$)~\cite{the_data}. 
The measurement shows an interesting structure around $\qinv$ 
(the magnitude of the proton momentum in the pair center-of-mass frame) 
of $70$~MeV/$c$, which statistically significantly deviates from the
expected unity.
NA49 has extensively studied possible systematic effects  
and none of these systematics can account for the structure.  
In this letter, we pursue possible physics explanations for the structure.
%While it is possible that the structure is due to 
%experimental systematics which have not been identified,
%we pursue possible physics explanations for the structure.

Similar structure has not been seen in other experiments for different 
collision systems or energies.  
Similar structures have been predicted in the tails of two-pion correlations 
\cite{makhlin99,pratt97} and 
have been ascribed to various effects, from hidden correlations
\cite{makhlin99} to a breakdown of the smoothness approximation
\cite{pratt97}.  Applying these predictions to two-proton correlations is 
difficult because of the differences between the two-proton and two-pion 
final state interactions.
%Given the differences between the two-proton and 
%two-pion final state interactions, the application of
%these predictions to two-proton correlations is difficult.  
Source imaging~\cite{HBT:bro97,HBT:bro98} 
is an ideal tool to separate effects due to two-proton 
final state interactions and wavefunction antisymmetrization 
from those due to the source function itself. 
%Source imaging~\cite{HBT:bro97,HBT:bro98} 
%is an ideal tool to remove the effects of the two-proton 
%final state interactions and wavefunction antisymmetrization 
%to access the source function directly.
Moreover, one can obtain the source function in a model-independent 
manner, i.e. without a Gaussian source assumption.

We have inverted the NA49 two-proton correlation function
and obtained a two-proton source
function that is exceptional in several ways.  
Our naive expectation was that
the source would appear somewhat Gaussian (as is the case for all 
two-pion and two-proton sources that have been obtained 
so far~\cite{HBT:bro97,HBT:bro98,HBT:bro98a,HBT:danielewicz98,pp:e895}). 
Instead, the source is consistent with a step-function with radius 
$\sim 10$~fm.
In some sense this is not a surprise as the
structure in the tail of the NA49 correlation function is reminiscent of the 
Fourier transform of a sharp object.  
In fact, the edge can be removed by smoothing out the structure. 
As if finding a step-function source is not strange enough, 
{\em this source cannot be explained within the standard 
Koonin-Pratt formalism}:
in the Koonin-Pratt formalism, the Fourier transform of a source must be  
positive everywhere while the Fourier transform of 
the NA49 source is not.  Thus, using imaging and this test 
of positivity, one may infer 
whether a given source is breaking {\em something}.
Knowing this, the task then becomes understanding what is broken.

The outline of this letter is as follows.  First, we describe the conventional 
expectations for sources in the Koonin-Pratt formalism.  Second, we detail  
why the Fourier transform of a source in the Koonin-Pratt formalism must 
be positive, thus giving us a test to see if a given source 
fits in this formalism.  
Then, we perform the inversion of the NA49 two-proton correlation function 
and show that it fails this test.  
We demonstrate that the structure in the tail of the correlation function 
is the origin of the failure.
Finally, we discuss physics that might be missing from the 
Koonin-Pratt formalism that could account for the unusual behavior of 
the source. 

%
%===========================================================
%
%\section*{Conventional Stuff}

We begin with our expectations for the source function in 
the Koonin-Pratt formalism~\cite{koonin_77,HBT:pra90,neq:dan92}.
A two-particle correlation function can be written as a convolution of 
a source function, $S_{q'}(\vec{r}\,)$, with the Wigner transform of the 
particle pair wavefunction, $g_{\vec{q}}(\vec{r},q')$, in the pair 
center-of-mass system (c.m.s.) \cite{neq:dan92}:
\begin{equation}
C(\vec{q}\,)=\int \dn{3}{r}\dntwopi{4}{q'} g_{\vec{q}}(\vec{r},q') 
S_{q'}(\vec{r}\,).
\label{eqn:ick}
\end{equation}
Here $\vec{q}=\frac{1}{2}(\vec{p}_1-\vec{p}_2)$ is the asymptotic particle 
momentum, $q'$ is the internal four-momentum 
(and need not correspond to on-shell particle momenta), 
%$P=p_1+p_2$ is the total momentum of the pair 
and $\vec{r}=\vec{r}_1-\vec{r}_2$ is the
separation of emission points.
Both $g_{\vec{q}}(\vec{r},q')$ and $q'$ are defined through
\begin{equation}
g_{\vec{q}}(\vec{r},q')=\int\dn{4}{\zeta} e^{i\zeta\cdot q'} 
\Phi_{\vec{q}}^*( r + \zeta/2) 
\Phi_{\vec{q}}( r - \zeta/2).
\label{eqn:gdef}
\end{equation}
Here $\Phi_{\vec{q}}(r)=\phi_{\vec{q}}(\vec{r}\,)e^{-iq_0 r_0}$ is the 
particle pair wavefunction.  Note that the different phases of the wavefunction 
in Eq.~\eqref{eqn:gdef} make $g_{\vec{q}}$ independent of the relative emission 
time.

In the Koonin-Pratt formalism, one makes two assumptions.  The first is that 
the source function only has a weak dependence on relative momentum 
corresponding to off-shell particles, 
i.e. $S_{q'}(\vec{r}\,)\approx S_{\vec{q}}(\vec{r}\,)$.
This is a {\em smoothness assumption}~\cite{pratt97,Weiner:1999th}.
With this, the integral over the internal momentum $q'$ in Eq.~\eqref{eqn:ick} 
can be performed and we obtain 
\begin{equation}
C(\vec{q}\,)=
\int d^3r \left| \phi_{\vec{q}}( \vec{r}\, )\right|^2 S_{\vec{q}}(\vec{r}\,).
\label{eqn:pk3D}
\end{equation}
%In the Koonin-Pratt formalism 
One further assumes that 
{\em the two-particle source is a convolution of single-particle sources} (i.e.
particles are emitted independently):
\begin{equation}
\begin{array}{rl}
\lefteqn{\displaystyle S_{\vec{q}}(\vec{r}\,) =}&\\ 
        &\displaystyle\int dt_1 dt_2 \dn{3}{R}\;
	D \left( \vec{R}+\frac{\vec{r}}{2},t_1; \vec{q} \right)\; 
	D \left( \vec{R}-\frac{\vec{r}}{2},t_2;-\vec{q} \right),
\end{array}
\label{eqn:defofsource}
\end{equation}
where $D$ is the normalized single-particle source in the c.m.s. 
and has the conventional interpretation as the normalized production rate of 
final-state particles.  The integral of $D$ over the particle emission 
times, $\int dt D(\vec{r},t;\vec{q}\,)$, is then the normalized 
phase-space distribution of particles after last collision (freeze-out) and
$S_{\vec{q}}(\vec{r}\,)$ is the 
probability density for a pair each with momentum $\vec{q}$ 
to be emitted a distance $\vec{r}$ apart in the c.m.s.
%Note, in \eqref{eqn:defofsource} the emission times of the protons are 
%integrated out.

In Eq.~\eqref{eqn:defofsource} we do not need to consider the contribution to 
the source from pairs with large relative momentum 
($|\vec{q}\,|=q\gtrsim q_{\rm cut} \approx 100$~MeV/$c$): 
the kernel cuts off the contribution from these pairs.  
This happens because the source varies slowly on the length scale of 
the oscillations in the kernel for 
$q\gtrsim q_{\rm cut}$, hence the integral in Eq.~\eqref{eqn:pk3D} averages to 
zero.  The fact that only low-$q$ pairs contribute to the correlation is
used to justify dropping the $q$ dependence from the single-particle sources
entirely \cite{HBT:pra90,gong91}.

The Fourier transform of Eq.~\eqref{eqn:defofsource} is 
\begin{eqnarray}
\lefteqn{\displaystyle\tilde{S}(\vec{k}\,) = 
\int \dn{3}{r} e^{i \vec{r}\cdot\vec{k}} S_{\vec{q}}(\vec{r}\,) =	
\int \dn{3}{R} \dn{3}{r} e^{i \vec{r}\cdot\vec{k}} \times \nonumber}&& \\
& & \int dt_1 \;D\left(\vec{R}+\frac{\vec{r}}{2},t_1; \vec{q}\,\right) 
    \int dt_2 \;D\left(\vec{R}-\frac{\vec{r}}{2},t_2;-\vec{q}\,\right) 
\nonumber \\
&\equiv&\;\tilde{D}(\vec{k},\vec{q}\,)\;{\tilde{D}}^*(\vec{k},-\vec{q}\,),
\end{eqnarray}
where $\tilde{D}$ is the Fourier transform of the time-integrated 
single-particle source.
{\em If we can neglect the $\vec{q}$ 
dependence of the source function} (or equivalently the momentum 
dependence of the single-particle sources in the c.m.s.),
then we find the following interesting result:
\begin{equation}
  \tilde{S}(\vec{k}\,) = \int \dn{3}{r} e^{i \vec{r}\cdot\vec{k}} S(\vec{r}\,)=
  |\tilde{D}(\vec{k}\,)|^2\geq 0.
  \label{eqn:3Dtest}
\end{equation}    
So, checking the Fourier transform of the source for positivity 
can serve as a test of the validity of the underlying assumptions.  

For pions, we note that Eq.~\eqref{eqn:3Dtest} is actually a restatement of the
theorem that the correlation function must always be larger than one 
(after the Coulomb correction). To see this, first realize that sources 
of indistinguishable particles must be symmetric under coordinate reversion:
$\vec{r}\rightarrow -\vec{r}$.  Using this, Eq.~\eqref{eqn:3Dtest} becomes
$\tilde{S}(\vec{k}\,) = \int \dn{3}{r} 
\cos(\vec{r}\cdot\vec{k}\,) S(\vec{r}\,)\geq 0$.  
Now, for like-pions Eq.~\eqref{eqn:pk3D} gives 
$ C(\vec{q})-1 = 
\int \dn{3}{r} \cos(2 \vec{r}\cdot\vec{q}\,) S(\vec{r}\,) =
\tilde{S}(2 \vec{q}\,)$
which must be positive.

We now turn to the practical issues of extracting information about the source
from the NA49 correlation function~\cite{the_data}.
The NA49 correlation function is averaged over the orientation of 
$\vec{q}$ so we integrate out the angular dependence of the correlation,
arriving at
\begin{equation}
  C(q)-1 = 4\pi \int dr\; r^2 K(q,r) S_{\vec{q}}(r).
  \label{eqn:pk1D}
\end{equation}    
Here $q=|\vec{q}\,|$ and $r=|\vec{r}\,|$.
The angle averaged kernel, 
$K(q,r)=\int\dn{}{(\cos\theta_{\vec{q}\vec{r}})}
|\phi_{\vec{q}}(\vec{r}\,)|^2 - 1$,  can be written as a sum over the 
two-proton partial waves:
\begin{equation}
  K(q,r) = \frac{1}{2} \sum_{js\ell\ell'}(2j+1) 
  \left(g_{js}^{\ell\ell'}(r)\right)^2-1.
  \label{eqn:kern}
\end{equation}
In what follows, we calculate the two-proton relative wavefunctions by 
solving the Schr\"odinger equation with the REID93~\cite{sto94} and 
Coulomb potentials for partial waves with $\ell\leq 2$
and use pure Coulomb waves for those with $\ell\geq 3$.  
Note that the results are not noticeably altered by using the full 
solutions to the Schr\"odinger equation for the higher partial 
waves due to the low relative momentum of the pair.
%~\cite{HBT:bro97,HBT:bro98}.

Since both Eqs.~\eqref{eqn:pk3D} and~\eqref{eqn:pk1D} are simple 
integral equations with non-singular kernels, they may be inverted
to obtain the source function.  We begin by discretizing 
either Eq.~\eqref{eqn:pk3D} or~\eqref{eqn:pk1D} to obtain the matrix equation
$C_i=\sum_j K_{ij} S_j$.  If the data were measured with infinite precision
we could invert this matrix equation to find $S_j=\sum_i (K^{-1})_{ij}C_i$.
Since we must account for the finite measurement errors, we proceed as
in~\cite{HBT:bro97,HBT:bro98,HBT:bro98a,HBT:danielewicz98} and find the
set of source points, $S_j$, that minimize the $\chi^2$.  Here, 
$\chi^2=\sum_i(C_i - \sum_j K_{ij} S_j)^2/\Delta^2C_i$.
This source is $S_j=\sum_i[(K^TBK)^{-1}K^TB]_{ji} (C_i-1)$ 
where $K^T$ is the transpose of the kernel matrix and $B$ is 
the inverse covariance matrix of the data
$B_{ij}=\delta_{ij}/\Delta^2C_i$.
The error on the source is the square-root of the diagonal elements of the
covariance matrix of the source, $\Delta^2S=(K^TBK)^{-1}$.

In this inversion process, 
all $q$ dependence of the correlation is ascribed to the kernel.  
If only the kernel has a $q$ dependence, then the imaging 
of Eq.~\eqref{eqn:pk3D} or~\eqref{eqn:pk1D} is unique.  
On the other hand, if the source also has a $q$ dependence, 
then the imaging gives the source averaged over $q$.
While only pairs with $q \lesssim q_{\rm cut}$ contribute to the correlation, 
this does raise the possibility that a strong $\vec{q}$ dependence
might cause trouble with the imaging.  
This possibility will be investigated in a future article~\cite{wiggle_theory}.

Averaging both sides of Eq.~\eqref{eqn:3Dtest} over angle, we obtain 
a condition for the angle averaged source (which we assume to have 
no $q$ dependence):
\begin{equation}
  \tilde{S}(k)\equiv 4\pi\int_0^\infty \dn{}{r} r^2 S(r) j_0(kr)\geq 0,
  \label{eqn:test}
\end{equation}
where $j_0(kr)$ is the $0^{\rm th}$ order spherical Bessel function.
The Fourier transform is a linear operation, so the errors propagate via
$\Delta\tilde{S}(k)=4\pi\int_0^\infty \dn{}{r} r^2 \Delta S(r) j_0(kr).$

%
%===========================================================
%
%\section*{The Analysis}

\begin{figure}
  \begin{center}
%    \epsfxsize=5.5in 
%    \epsffile{fig1.eps}
%    \psfig{figure=fig1.eps,height=3in,angle=-90}
    \includegraphics[width=\textwidth]{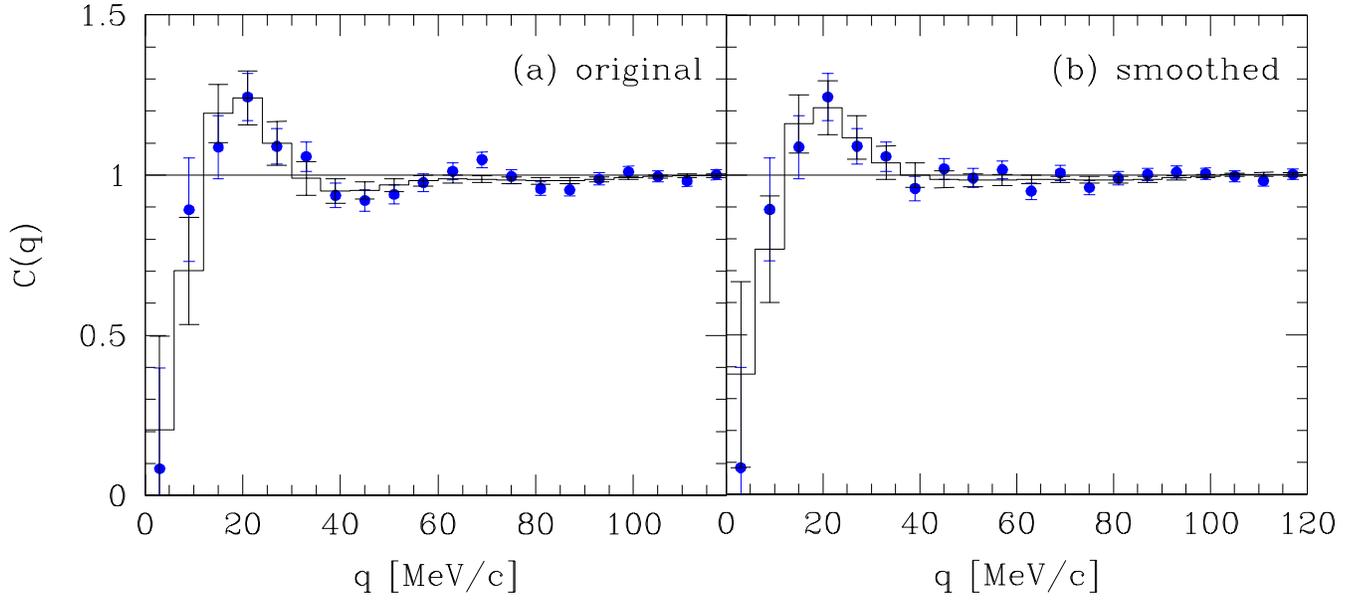}
  \end{center}
  \caption[]{
	(a) Correlation function from ref.~\cite{the_data} 
	(filled circles, small caps on error bars) 
	and the restored correlation function (solid 
	histogram, large caps on error bars).
	(b) Same as (a) but the structure between $36<q<102$~MeV/$c$ is 
	replaced by a flat correlation with 
	experimental error and realistic statistical scatter (filled circles, 
	small caps on error bars), 
	and the restored correlation function (solid
	histogram, large caps on error bars).}
  \label{fig:corr}
\end{figure}

\begin{figure}
  \begin{center}
%    \epsfxsize=5.5in 
%    \epsffile{fig2.eps}
%    \psfig{figure=fig2.eps}
    \includegraphics[width=\textwidth]{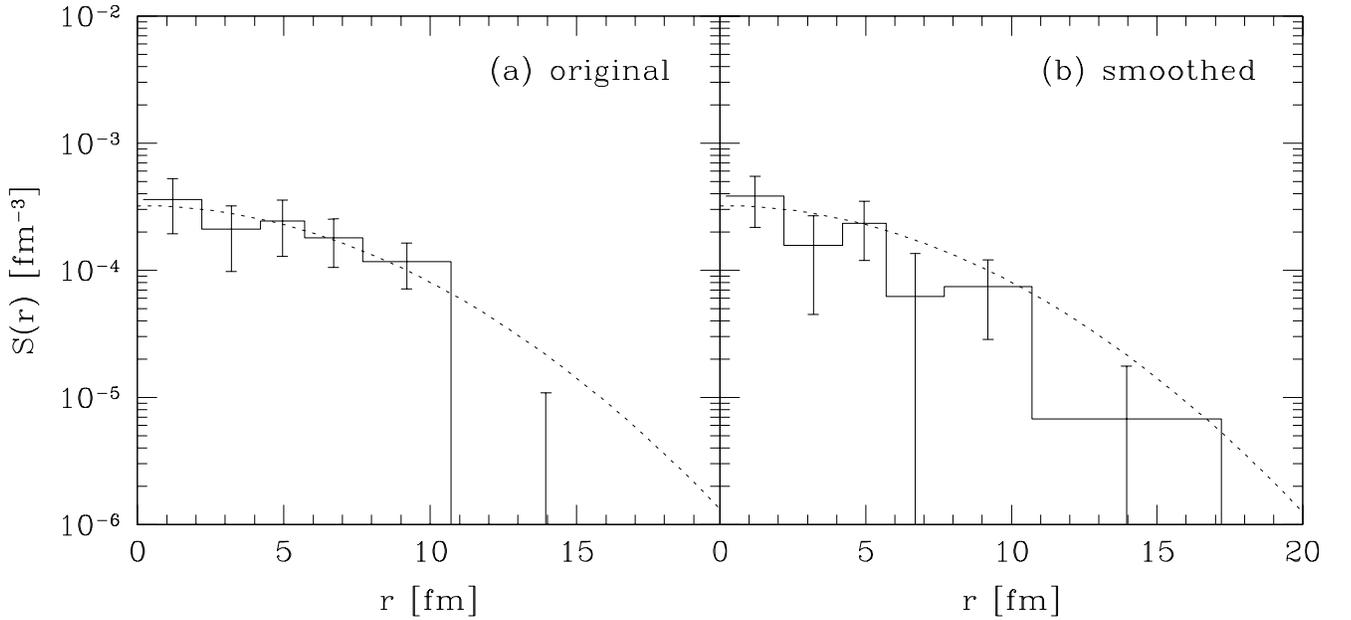}
  \end{center}
  \caption[]{(a) Source function obtained by imaging the correlation data 
	(solid histogram) and NA49's best-fit Gaussian source (dashed curve).  
	(b) Source function obtained by imaging the ``smoothed'' 
	correlation function (solid histogram) also with NA49's best-fit 
	Gaussian source (dashed curve).}
  \label{fig:source}
\end{figure}

We now turn to the NA49 two-proton correlation data~\cite{the_data}.  
In Fig.~\ref{fig:corr}a, we reproduce the data with the filled circles.
Note the oscillation at $q\sim$70~MeV/$c$ and maybe at $\sim$100~MeV/$c$.  
These oscillations are responsible for all the interesting behavior 
we discuss below.
Figure~\ref{fig:source}a shows the source function obtained by 
imaging the correlation function in Fig.~\ref{fig:corr}a. 
No constraints have been imposed in the imaging.  
Note that the source is nearly flat up to $r\simeq 10$~fm, 
after which the source drops rapidly to zero in the bin centered at 14~fm.  
The source remains several orders of magnitude smaller
than the main part of the source out to the edge of the image at 
$r \approx 65$~fm.  For comparison, we plot NA49's best-fit source 
in the dashed curve in Fig. 2a.  NA49 fit a Gaussian 
single-particle source to the $q<48$~MeV/c points of the correlation 
function and found a best-fit Gaussian radius of 
$3.85$~fm.  Examining Fig. 2a., the best-fit Gaussian source
describes the $r\leq 10$~fm points and fails beyond that.
As a side benefit of imaging, we can extract the integral 
of the source over the imaged region.  We find 
$4\pi \int_0^{12{\rm fm}} dr\; r^2 S(r) = 0.78 \pm 0.55$.
We should mention that the constant weak decay correction applied in 
ref.~\cite{the_data} does increase the magnitude, and hence the integral, 
of the source over the uncorrected one, but the correction does not 
change the structure of the oscillation.  

As a consistency check, we uninverted the source into the correlation 
function which is shown as the solid histogram in Fig.~\ref{fig:corr}a. 
%The resulting correlation function is consistent with the original 
%but, due to the loss of information in the inversion-then-uninversion 
%process, the error on the correlation has increased and the structure 
%in the tail of the correlation has been washed out.
The resulting correlation function is consistent with the original 
but the structure in the tail of the correlation has been washed out.
This is easily understood as an effect of the loss of 
information in going from a correlation function 
with $\sim 20$ independent data points to a source with only 7 points.

\begin{figure}
  \begin{center}
%    \epsfxsize=5.5in 
%    \epsffile{fig3.eps}
%    \psfig{fig3.eps}
    \includegraphics[width=\textwidth]{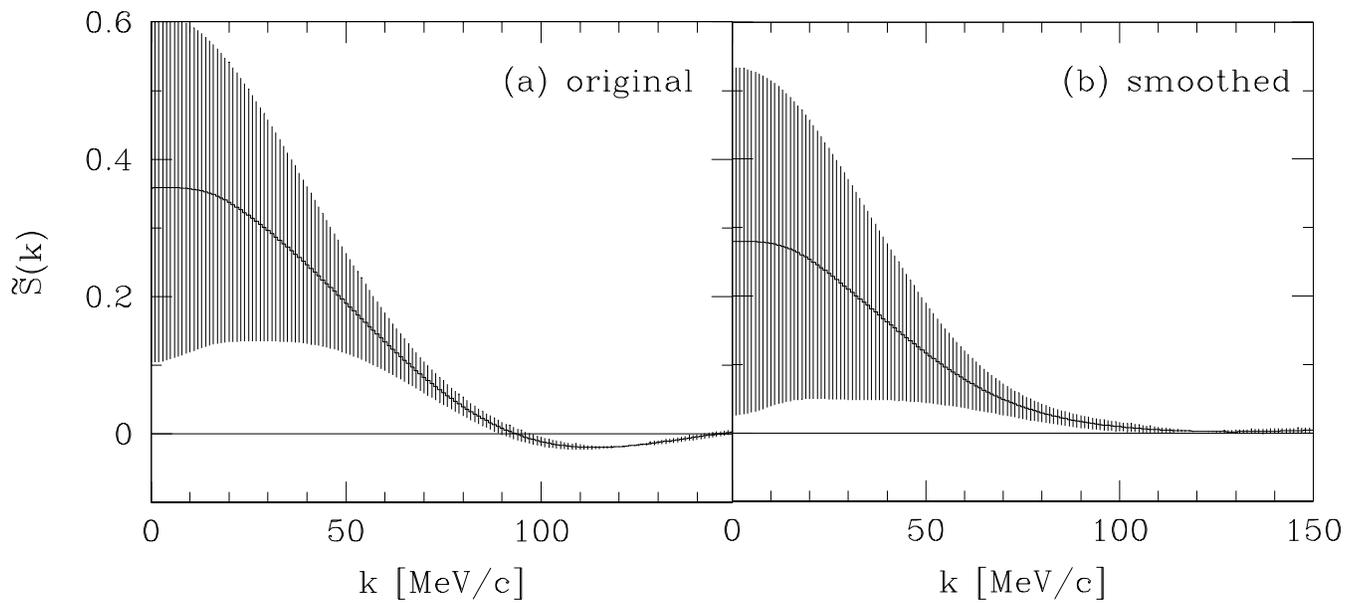}
  \end{center}
  \caption[]{
%	\debug{I propose to label the plots with $\tilde{S}(k)$ vs $k$}{}\\
	(a) Fourier transform of the source function in 
	Fig.~\ref{fig:source}a.  Notice the dip below zero 
	in the range $100<k<150$~MeV/$c$.  
	(b) Fourier transform of the source function in 
	Fig.~\ref{fig:source}b.  Notice that this function 
	is everywhere positive.  In both panels, the vertical lines represent 
	the error band of the calculation.}
  \label{fig:test}
\end{figure}

Now we test whether the imaged source fails the Fourier transform 
test discussed above.  We must take care in doing the integral 
in Eq.~\eqref{eqn:test} as simply discretizing it introduces numerical error 
at high $k$, exactly where a failure of the Fourier transform test will occur.
Instead, we use an interpolation procedure from ref.~\cite{numrec} to 
evaluate Eq.~\eqref{eqn:test} with the $S(r)$ in Fig.~\ref{fig:source}a.
The result of this integration is shown in Fig.~\ref{fig:test}a.
$\tilde{S}(k)$ goes negative in the range 100--150~MeV/$c$.

For comparison, in Fig.~\ref{fig:corr}b. we replace the structure in the 
tail of the correlation with statistical noise.  
To be exact, we replace the measured correlation function in 
$36<q<102$~MeV/$c$ with a correlation consistent with one, 
but containing the original errors and statistical scatter.
Applying the imaging technique, we obtain the source function shown in 
Fig.~\ref{fig:source}b.  For comparison, we plot the best-fit Gaussian
source along with the ``smoothed'' source in Fig. 2b and the two sources 
are consistent.  Furthermore, not only does this modified source appear 
less flat and somewhat Gaussian
but its Fourier transform is positive as shown in Fig.~\ref{fig:test}b.
Thus we have demonstrated that the structure in the tail of the correlation 
is the cause of the failure shown in Fig.~\ref{fig:test}a.  
This assertion should not come as a surprise as the behavior of the 
correlation at the scale $q$ 
is responsible for source structure on the scale $r \sim \pi\hbar c/q$.

At this point we comment on the model comparisons by NA49 using
different single-particle and two-particle sources \cite{the_data}.
%In attempting to explain the structure, NA49 has tried different
%single-particle and two-particle sources \cite{the_data}.
First, NA49 compared their data to correlations generated from VENUS and RQMD 
simulations.  Both models generate the final freeze-out distribution
of protons (equivalent to our $D(\vec{r},t;\vec{q})$) and these distributions
are input to a correlation afterburner that performs the integrals in 
Eqs.~\eqref{eqn:defofsource} and~\eqref{eqn:pk3D}.  Thus, by construction 
these correlations pass the Fourier transform test and cannot describe the
tail of the correlation data.  NA49 also tried a hard edged two-proton
source but could not reproduce the observed structure in the correlation
without abandoning the independent source assumption.

%
%===========================================================
%
%\section*{Conclusion}
 
%We are now left with a daunting task:  
%if the structure in the tail of the NA49 correlation function is verified, 
%then we must conclude that something is wrong with either the imaging, 
%the assumptions in the Koonin-Pratt formalism or both.  
We are now left with a dilemma:  
something is wrong with either the imaging, 
the assumptions in the Koonin-Pratt formalism or both.
While experimental confirmation of this new data is needed,
we suggest three possible physics sources of the problem 
(as we have indicated in italic text in this letter):  
(1) a strong low-$q$ dependence of the source that is not imaged, 
(2) the source not being a convolution of single-particle sources,
or (3) a break-down of the smoothness assumption.

In the case that there is a strong low-$q$ dependence 
in the NA49 source function, such a $q$ dependence cannot be reconstructed 
by the imaging.  In this case, the Fourier transform test is inappropriate
as the imaged source does not reflect the true NA49 source.  
%A strong low-$q$ dependence of the source might result from strong 
%position-momentum correlations in the single-particle sources.  
%However, model studies do not indicate this~\cite{newsergei,gong91}.
One would think that a strong low-$q$ dependence of the source 
might result from strong position-momentum correlations in the 
single-particle sources.  In ref.~\cite{newsergei}, the authors deliberately 
increased the degree of position-momentum correlation
by aligning the proton transverse momentum with the transverse
position of the proton at freeze-out in RQMD.  Even after
this dramatic increase in the position-momentum correlation, they were
able to image the source function reliably.

In the second case, the NA49 two-proton source is not a convolution
of two independent sources.  In this case, both imaging and the Fourier 
test are appropriate.  However, interpreting the source function becomes 
difficult as we cannot model it simply with a transport model.
Makhlin and Surdutovich~\cite{makhlin99} suggest that this possibility 
could occur if there are hidden correlations in the system.  Such hidden 
correlations would arise from having three or more dynamically or statistically 
correlated particles in the final state while only observing two of them.  

Finally, in the event that the smoothness assumption~\cite{pratt97}
is no longer valid, the entire formalism breaks down as we can no longer 
perform the integral over the internal four-momentum in Eq.~\eqref{eqn:ick}.
%factorize the two-particle relative wavefunction in Eq.~\eqref{eqn:pk3D} 
%from the source.  
This is the worst possible case from our standpoint as 
%imaging can no longer be applied.
neither Eq.~\eqref{eqn:pk3D} nor Eq.~\eqref{eqn:pk1D} is valid and we 
cannot apply our imaging procedure.
Pratt~\cite{pratt97} argues that this 
could be caused by strong off-shell effects and these effects would be 
magnified by strong position-momentum correlations in the source
and by the short-range interaction between the protons.  Pratt
comments that these effects are larger in small systems, e.g. with  
single-particle source size $R\sim 1$~fm, 
much smaller than the source imaged here.

In conclusion, we have imaged the recently measured NA49 two-proton 
correlation function and found an extraordinary two-proton source function. 
This source function appears to be a step function and we show that the 
steep fall-off in the source is due to the structure in the tail of the 
NA49 correlation function.
Furthermore, we show that this source does not fit into 
the Koonin-Pratt formalism. 
%%Provided the data is verified, identifying what is responsible
%%for the unusual structure in the source will require further insight.
%\debug{Identifying what is responsible
%for the unusual structure in the source will require further insight.}
%Nevertheless, imaging and the application of the Fourier transform test to the 
%reconstructed sources are valuable tools.
While we have suggested several possible physics explanations for the structure 
in the two-proton correlation, its origin remains a puzzle.

%
%===========================================================
%
\section*{Acknowledgements}
The authors gratefully acknowledge discussions with 
George Bertsch, Richard Lednicky, Assum Parre\~no, Sergei Panitkin, John 
Cramer, and Nu Xu.   
This work was supported by the U.S. Department of Energy under grants
DOE-ER-40561 and DE-AC03-76SF00098 and by the National Science Foundation 
under grant PHY-96-05207.

%
%----------------------------------------------------------------------
%   Bibliography
%

%----------------------------------------------------------------------
%   Figures
%
\onecolumn
%\printfigures

\newpage

%
%-----------------------------------------------------------
\end{document}